\pdfoutput=1
\documentclass[aps,twocolumn,superscriptaddress]{revtex4-2}
\usepackage[colorlinks=true,citecolor=blue,urlcolor=blue,linkcolor=blue,pdfstartview=FitH,bookmarksopen]{hyperref}
\usepackage{amsmath,amssymb}
\usepackage{bm}
\usepackage{comment}
\usepackage{graphicx}
\usepackage[dvipsnames]{xcolor}
\usepackage{custom}

\begin{document}

\title{Relativistic Guiding-Center Motion: Action Principle,\\ Kinetic Theory and Hydrodynamics}

\author{Dam Thanh Son}
\affiliation{Kadanoff Center for Theoretical Physics, University of Chicago, Chicago, Illinois 60637, USA}
\author{Mikhail Stephanov}
\affiliation{Kadanoff Center for Theoretical Physics, University of Chicago, Chicago, Illinois 60637, USA}
\affiliation{Department of Physics, University of Illinois, Chicago, Illinois 60607, USA}

\begin{abstract}
We treat the guiding-center dynamics in a varying external Maxwell field using a relativistically covariant action principle which reproduces the known Vandervoort expression for the drift velocity and extends it to curved spacetime. We derive the corresponding kinetic theory and ideal hydrodynamic theory. In contrast to conventional five-equation hydrodynamics, the guiding-center hydrodynamics needs only three equations due to a constraint on the motion across magnetic field. We argue that such a hydrodynamics is applicable to strongly coupled plasmas where kinetic theory fails.
\end{abstract}

\date{May 2024, revised July 2024}

\maketitle

\emph{Introduction}---About 50 years ago, Vitaly Ginzburg identified matter in ultrahigh magnetic fields as one of the most important and interesting problems in physics~\cite{Ginzburg-book}.  This problem remains current
to this day.
In astrophysics, magnetic fields exceeding $10^{15}~\text{G}$ have been found on the surface of a type of neutron star called a magnetar~\cite{Kaspi:2017fwg,Olausen:2013bpa}.  In peripheral ultrarelativistic heavy-ion collisions, the magnetic field created by the colliding nuclei may exceed $10^{19}~\text{G}$~\cite{Skokov:2009qp,Voronyuk:2011jd} and leave measurable effects on the collisions~\cite{STAR:2023jdd}.  In condensed matter physics, magnetic fields available in the lab  ($\sim\! 10^5~\text{G}$) can already be considered strong at low densities of charge carriers and temperatures, and lead to nontrivial phenomena like the integer and fractional quantum Hall effects~\cite{vonKlitzing:1980pdk,Tsui:1982yy}.

It is known that  properties of matter change dramatically in high magnetic fields~\cite{Lai:2000at}.  Ordinary atoms become deformed when the magnetic field $B$ reaches $m_e^2c e^3/\hbar^3\approx 2.35\times10^9~\text{G}$ and, when $B$ exceeds $m_e^2c^3/(e\hbar)\approx 4.41\times10^{13}~\text{G}$, the cyclotron motion of electrons is relativistic. Electron motion at high $B$ can be decomposed into a fast cyclotron motion and a slower motion of the orbit's guiding center.  At very high $B$, the cyclotron motion is quantum and, in the most extreme case, frozen onto the lowest Landau level. The motion of the guiding center, on the other hand, is (semi)classical and is a textbook problem of plasma physics. It is well known that the guiding center moves mostly along the magnetic field line with a slow ``drift'' off that line when the field is inhomogeneous~\cite{Jackson-EM}. 
This drift motion is important in various physical problems including cosmic-ray transport~\cite{Jokipii:1977} and thermonuclear plasma confinement~\cite{Freidberg-book}.
A plasma of electrons in a high $B$ field can be thought of as a plasma of guiding centers, and finding a description of the collective dynamics of these guiding centers is crucial for understanding the behavior of matter in very high magnetic fields.  Guiding-center dynamics is used in simulations of astrophysical plasmas (see, e.g., Refs.~\cite{Gordovskyy:2014,Threlfall:2015,Ripperda:2017ifb,Bacchini:2020cze,Mignone:2023}).

The goal of this Letter is to develop a fully relativistic, manifestly Lorentz covariant description of the guiding-center motion, and from that derive the kinetic theory of the guiding-center plasma, which, in turn, underlies a hydrodynamic description of the plasma at sufficiently long length and time scales.

\emph{Relativistic motion of the guiding center}---%
In the special relativistic setting, the problem of guiding-center motion has been 
solved since the 1960s~\cite{Vandervoort:1960,Northrop-book} and revisited recently in Refs.~\cite{Beklemishev:1999,Beklemishev:2004,Trent:2023tnu,Trent:2024pii}. Here we provide an alternative, simpler, and manifestly Lorentz covariant description of the motion based on an action principle. This formulation is easy to generalize to curved spacetime and will turn out to be particularly apt for our subsequent derivation of the kinetic and hydrodynamic theories.

We assume the presence of a strong electromagnetic field $F_{\mu\nu}$ which is magnetically dominated, i.e., $ F_{\mu\nu}F^{\mu\nu}=2( \bm B^2-\bm E^2)>0$.  At any given spacetime point there exists a frame where $\bm E\!\parallel\!\bm B$.
We denote by
$\FMs$ and $\FEs$ the magnitudes of the electric and magnetic fields in that frame ($\FEs<0$ if the fields are antiparallel).  In terms of the Lorentz invariants, $F_{\mu\nu}F^{\mu\nu}=2(B_*^2-E_*^2)$, $F_{\mu\nu}\tilde F^{\mu\nu}=-4E_*B_*$~\footnote{We use the convention of Refs.~\cite{Jackson-EM,LL2}: The metric signature is $(+,-,-,-)$, $A^\mu=(\phi,\mathbf{A})$, $\tilde F^{\mu\nu}=\frac12\epsilon^{\mu\nu\alpha\beta}F_{\alpha\beta}$, where $\epsilon^{\mu\nu\alpha\beta}$ is the Levi-Civita tensor, $\epsilon^{0123}=+1$.}.

We consider the motion of an electron in this field and, to simplify further order-of-magnitude estimates, assume that the motion is relativistic: $v\sim 1$.  Let $p_\perp$ be the magnitude of the electron's momentum in the plane perpendicular the magnetic field. Then, the cyclotron radius is $a=p_\perp/(eB_*)$.  We assume $a$ is much smaller than the scale $L$ on which the magnetic field varies: $a\ll L$. In that case, the motion of the electron is a superposition of a fast cyclotron motion and the smoother motion of the guiding center. We can then expand in the gradients of the magnetic field, with each additional gradient being suppressed by $\mathcal O(a/L)$.

Furthermore, we assume that the electric field $E_*$ is small enough so that the longitudinal momentum that the electron obtains while accelerating in that field during a time $\sim\!\! L$ does not exceed $p_\perp$: $eE_*L \lesssim p_\perp$.  It follows that $E_*/B_* \lesssim a/L \ll 1$.  Therefore, we assume that the longitudinal electric field is suppressed by the small parameter $a/L\ll1$, $E_*/B_* = O(a/L)$.

In the frame where $\bm E\!\parallel\! \bm B$, one can separate $F_{\mu\nu}$ into a purely magnetic part $B_{\mu\nu}$ and purely electric part $E_{\mu\nu}$.
Boosting back to the original frame, we find
\begin{equation}
  \label{eq:BEvsF}
  \FMu_{\mu\nu} = \frac{F_{\mu\nu}-\epsilon_*\tilde
    F_{\mu\nu}}{1+\epsilon_*^2},
  \quad
  \FEu_{\mu\nu} = \epsilon_*\frac{\tilde F_{\mu\nu}+\epsilon_*
    F_{\mu\nu}}{1+\epsilon_*^2},
\end{equation}
where $\epsilon_*\equiv E_*/B_*$. 

The equations governing the motion of the guiding center were first derived in Ref.~\cite{Vandervoort:1960} (see also Ref.~\cite{Northrop-book}) in explicit component form where Lorentz invariance is hidden. 
We show that these equations can be obtained straightforwardly from the manifestly Lorentz invariant least action principle for a relativistic particle with an additional constraint.  Namely, let the trajectory of the particle in phase space be $(x^\mu(\tau), p_\mu(\tau))$, where $\tau$ is an affine parameter along the worldline.
Consider the following Lagrangian:
\begin{equation}\label{eq:Lagrangian}
  L = - \dot x\cdot (p + A) + \frac\alpha 2 (p^2 -\tilde m^2) + \lambda_\mu \FMd^{\mu\nu} p_\nu ,
\end{equation}
where $\dot x\equiv dx/d\tau$.
For convenience, we have absorbed the charge of the particle into the vector potential $A$.
The second term implements, via a Lagrange multiplier $\alpha$, the on-shell condition $p^2=\tilde m^2$. The last term implements, via Lagrange multipliers $\lambda_\mu$, the constraint  $B^{\mu\nu}p_\nu=0$. In the $\bm E\!\parallel\!\bm B$ frame, this condition becomes $\bm B\times\bm p =0$.
This reflects the result of averaging of the momentum of the particle perpendicular to $\bm B$ over the fast orbital motion: $\overline{\bm p_\perp}=0$~\footnote{One can integrate out $p$ and $\lambda$ and arrive at a Lagrangian in terms of the guiding center's coordinate-space worldline~\cite{SM}.  The phase-space description is, however, more convenient for deriving the kinetic equation.}. 

Another effect of the orbital motion is that $\tilde m$ in Eq.~(\ref{eq:Lagrangian}) is not the original particle's rest energy.  Indeed, the on-shell condition for the original particle reads $p_0^2=p_\parallel^2+\overline{p_\perp^2}+m^2$, therefore $\tilde m^2 = m^2+\overline{p_\perp^2}$. Recalling that there is an adiabatic invariant equal to the particle's orbital momentum,
\begin{equation}\label{eq:J-adiabatic}
J=ap_\perp=p_\perp^2/\FMs ,
\end{equation}
we can then write
\begin{equation}\label{eq:mt=mJB}
    \tilde m^2 = m^2 + J\FMs \,.
\end{equation}

In the quantum case, $J=\hbar(2n+1+gs_z)$ where $n$ labels the Landau levels, $s_z=-s, -s+1,\ldots, s$ where $s$ is the spin of the particle, and $g$ is its $g$-factor.

Varying the action $S\equiv \int L d\tau$ with respect to $p_\mu$, $x^\mu$, $\alpha$, and $\lambda_\mu$ we obtain the equations of motion
\begin{subequations}\label{eq:eoms}
\begin{align}
  \dot x^\mu &= \alpha p^\mu +\lambda_\alpha \FMd^{\alpha\mu} 
  \equiv \alpha p^\mu
  + v_\mathrm{D}^\mu ,\label{dotx} \\
  \dot p_\mu &= 
  F_{\mu\nu}\dot x^\nu + \frac \alpha2 J
               \partial_\mu \FMs - \lambda_\alpha p_\beta \partial_\mu
               \FMd^{\alpha\beta} ,
               \label{eq:pdot}
 \end{align}     
\end{subequations}
together with two constraints $p^2=\tilde m^2$ and $\FMd^{\mu\nu}
p_\nu=0$.  From Eq.~(\ref{dotx}) one sees that $\alpha$ and $\lambda_\alpha$ depend on the choice of the affine parameter $\tau$.  For example, if $B_{\mu\nu}$ is purely spatial, choosing $\tau=t$ means that $\alpha=1/p_0$.

While the momentum of the guiding center always points in the direction of the magnetic field due to the constraint $\FMd^{\mu\nu}p_\nu=0$, the particle does not move strictly along the magnetic field lines: It also slowly drifts in an orthogonal direction with a velocity given by the spatial components of $  v_\mathrm{D}^\mu \equiv \lambda_\alpha \FMd^{\alpha\mu}\,
$. Substituting Eq.~\eqref{dotx} into Eq.~\eqref{eq:pdot} and multiplying by $\FMd^{\lambda\mu}$, we find
\begin{equation}\label{eq:vD-dB}
  v^\mu_\mathrm{D} = \alpha\FMs^{-2}\!\left(
    (p\cdot \partial) \FMd^{\mu\nu}p_\nu+
    \frac{J}{2} \FMd^{\mu\nu} \partial_\nu \FMs
    \right),
\end{equation}
where we
dropped terms higher order in gradients of the magnetic field.
Note that, since $v_\mathrm{D}$, and thus $\lambda$, is first order in gradients of $B$, the last term in Eq.~\eqref{eq:pdot} is
second order in the gradients.

Equation~(\ref{eq:vD-dB}) compactly and covariantly summarizes the results of Refs.~\cite{Vandervoort:1960,Northrop-book}. The first and the second terms reproduce the drifts due to the change of the magnetic field direction (curvature drift) and magnitude (gradient drift), respectively. Additional effects due to time derivatives of the fields, dictated by Lorentz invariance, are also included~\cite{SM}.

The covariant generalization of Eqs.~\eqref{eq:Lagrangian}, \eqref{eq:eoms}, and~\eqref{eq:vD-dB} to curved spacetime is straightforward; see Eqs.~\eqref{eq:Lagrangian-curved}--\eqref{eq:vD-dB-curved} in the Appendix. 
In particular, the spacetime derivatives in Eq.~\eqref{eq:vD-dB} are simply replaced by covariant derivatives.

\emph{Kinetic equation}---Using the equation of motion for a single particle derived above, we now study the kinetic theory of a collection of charged particles moving in a large external, almost crossed  ($\epsilon_*\ll1$), field.  To be able to replace the particle motion by that of the guiding centers, the cyclotron radius must be shorter than the mean free path.  For example, for a relativistic electron-positron plasma with $T\gtrsim m_e$, the cyclotron radius is $T/(eB)$, while the mean free path is (up to a log) $1/(e^4T)$.  The kinetic theory we are going to derive works when $B\gg e^{3}T^2$.  

The relativistic kinetic equation of a gas of guiding centers is
given by
\begin{equation}\label{eq:L=C}
\mathcal L[f]\equiv
\dot x\cdot\frac{\partial f}{\partial x}
  + \dot p\cdot\frac{\partial f}{\partial p}
=\mathcal C[f]\,,
\end{equation}
where the Liouville operator $\mathcal L[f]$ represents the rate of change (per unit worldline ``time'') of the occupation number $f(x,p)$ along the particle worldline 
determined by Eqs.~\eqref{eq:eoms}. The occupation number is changed by collisions; therefore, the rate is equal to the collision integral
$\mathcal C[f]$.

To describe transport in the gas, we need to know also the invariant phase-space integration measure, i.e., the number of single-particle states in a given phase-space volume at a given time in the lab frame. The number of states is time independent if the phase-space volume evolves according to the equations of motion. This statement can be expressed in Lorentz covariant form by introducing the conserved phase-space 8-current $J^M$ given by~\cite{Stewart-book}
\begin{equation}
  \label{eq:JM}
  J^M = 
  (\dot x^\mu,\dot p_\nu)
  W \,,
\end{equation}
i.e., by the product of the phase-space ``velocity'' and the ``density'' $W$~\footnote{We need to extend the velocities $\dot x^\mu$, $\dot p_\nu$ to the whole eight-dimensional phase space.  We can do that by simply declaring Eqs.~(\ref{eq:eoms}) and (\ref{eq:vD-dB}) to be valid everywhere, even outside the constraint hypersurface}.

The flux of $J^M$ through a spacelike hypersurface (e.g., a constant-time hypersurface in a given frame with 4-velocity $n$ with surface element $d^7\Sigma_M=(n_\mu,0)d^3xd^4p$) gives the number of states on that hypersurface:
\begin{equation}
  \label{eq:dGamma}
  d^7\Gamma = J^Md^7\Sigma_M = (n\cdot\dot x)Wd^3xd^4p.
\end{equation}
The condition for the conservation of current $\d_M J^M=0$, together with
the constraints on $p$ that limit the support of $W$ to the intersection of the mass-shell hyperboloid $p^2=\mt^2$, $p_0>0$ and the hyperplane $B^{\mu\nu}p_\nu=0$, can be solved to yield~\cite{SM}
\begin{equation}
  \label{eq:W}
    W = \frac{\FMs+B_*^{-1} p_\mu\d_\nu\FMd^{\mu\nu}}{2\pi^2\alpha} \theta(p_0)
  \delta(p^2-\tilde m^2)\delta^2(\Delta^{\mu\nu}p_\nu),
\end{equation}
where we have defined the projector $\Delta^\lambda_\nu$
as
\begin{equation}\label{eq:projectors}
  \Delta^\mu_\nu = B_*^{-2} B^{\mu\alpha} B_{\nu\alpha}, \qquad
  \tilde\Delta^\mu_\nu = \delta^\mu_\nu - \Delta^\mu_\nu.
\end{equation}
For a magnetic field along the $z$ axis, $\Delta^\mu_\nu$ projects onto the $(x,y)$ plane, while the complement $\tilde\Delta^\mu_\nu$, onto the $(t,z)$ hyperplane. 

Note that the equation $\partial_M J^M=0$ determines 
$W$ only up to an overall  constant. This constant is chosen in Eq.~\eqref{eq:W} to match the Landau-level degeneracy.  

\emph{Hydrodynamics}---We can now use the guiding-center kinetic theory to describe the transport in the hydrodynamic regime.  For simplicity, we consider a single-component plasma neutralized by an inert background charge density.  The electromagnetic field is treated mostly as a fixed background, but can be made dynamical by supplementing the hydrodynamic equations with Maxwell's equations.

We first summarize the main result.  To lowest order in gradients, the transport is described by the ideal hydrodynamic equations expressing three conservation laws,
\begin{subequations}
\label{eq:dJdTEJ}
    \begin{align}
  \d_\mu \Jt^\mu &= 0 \label{eq:dJt=0},\\
\tilde\Delta^\lambda_\nu \d_\mu T^{\mu}_\lambda &= E_{\nu\lambda} \Jt^\lambda . \label{eq:projected-dmuTmunu}
\end{align}
\end{subequations}
The presence of the projector $\tilde\Delta$ means that Eq.~(\ref{eq:projected-dmuTmunu}) represents only two independent equations, not four.  These conservation laws are supplemented by the constitutive relations that express current $\Jt^\mu$ and  stress tensor $T^{\mu\nu}$ in terms of three independent hydrodynamic variables: the local temperature $T(x)$, local chemical potential $\mu(x)$, and local velocity $u^\mu(x)$.  The velocity is subject to the constraints $u\cdot u=1$, $B_{\mu\nu}(x)u^\nu(x)=0$ and hence represents only 1 degree of freedom.  

The constitutive relations necessary to close the system of hydrodynamic equations read 
\begin{subequations}\label{eq:JT=}
\begin{align}
   \Jt^\mu 
    &= n u^\mu\,,
   \label{eq:J=}
   \\
   T^{\mu\nu} &= (\epsilon + P) u^\mu u^\nu - P\tilde\Delta^{\mu\nu}
      - P_\perp \Delta^{\mu\nu},
   \label{eq:T=}
\end{align}
\end{subequations}
where $n$, $\epsilon$, $P$, and $P_\perp$ are functions of $T$ and $\mu$ and can all be recovered from a single thermodynamic potential $P=P(T,\mu,B_*)$ by taking derivatives or various Legendre transforms,
\begin{equation}\label{thermo-rels}
  n =\frac{\d P}{\d\mu}\,, ~
  \epsilon = T\frac{\d P}{\d T}
  + \mu \frac{\d P}{\d\mu} - P\,, ~
  P_\perp = P - B_* \frac{\d P}{\d B_*}\,.
\end{equation}
The anisotropy of the pressure tensor due to the magnetic field seen in Eq.\eqref{eq:T=} was discussed, e.g., in Refs.~\cite{Chew:1956,Tenbarge:2008,Hernandez:2017mch,Lingam:2020,Hattori:2022hyo}.

\emph{Derivation of hydrodynamics}---Multiplying Eq.~(\ref{eq:L=C}) by $W$ and $Wp_\nu$ and integrating over $p$, using the fact that the collision term $\mathcal C[f]$ conserves the particle number (charge) and 4-momentum, i.e. \footnote{We have chosen the charge of particles to be 1. If there is more than one type of particle, one needs to sum over particle types, labeled by $a$. For example, conservation of charge becomes $\sum_a \int_p W q_a C[f_a]=0$ where $q_a$ is the charge of the particles of type $a$. Similarly, if the values of adiabatic invariant $J$ vary, they have to be summed over too, e.g., in the second equation of \eqref{eq:Force}.},
\begin{equation}
 \int_p\, W \{1,p_\nu\} \mathcal C[f] = 0\,,
\end{equation}
where $\int_p\equiv \int\!d^4p$, 
we find the conservation laws
\begin{align}\label{eq:dJT=0F}
  \d_\mu \{\Jxdot^\mu, \Tt^\mu_{~\,\nu}\} = \{0,F_\nu\} \,,
\end{align}
where 
\begin{align}
   \{\Jxdot^\mu, \Tt^\mu_{~\,\nu},F_\nu\} = \int_p W  \{\dot x^\mu ,\dot x^\mu p_\nu, \dot p_\nu\}f\,.
  \label{eq:TmunuF}
\end{align}
As written, $\tilde T^{\mu\nu}$ is not symmetric, but by replacing $\dot x^\mu$ with $\alpha p^\mu$ one makes an error suppressed by one power of gradients.  Therefore, to lowest order in gradients, one can write
\begin{equation}
  \label{eq:Tmunu}
  \Tt^{\mu\nu} = \int_p W\alpha\, p^\mu p^\nu f .
\end{equation}
By using Eq.~(\ref{eq:pdot}) and ignoring higher-order contributions, the force term in Eq.~(\ref{eq:TmunuF}) can be expressed as
\begin{equation}
  \label{eq:Force}
  F_\nu 
= F_{\nu\lambda} \Jxdot^\lambda - M \partial_\nu\FMs , ~~ 
  M\equiv -\!\int_p W\frac{\alpha J}2 f 
\end{equation}
where $M$ is the magnetic moment density ($-\alpha J/2$ per particle).

We will call $N^\mu$
in Eq.~\eqref{eq:TmunuF} the ``transport current'' to distinguish it from the full current, to be introduced later.  It consists of two parts:
$\Jxdot^\mu=\Jt^\mu+\JD ^\mu$, where the longitudinal current 
\begin{equation}
 \label{eq:Jpar-f-app}
  \Jt^\mu \equiv \int_p W\alpha p^\mu  f 
\end{equation}
flows along the magnetic field lines ($\FMu_{\mu\nu}\Jt^\nu=0$), while the drift current
\begin{equation}
 \label{eq:JD-Tmunu}
   \JD^\mu \equiv \int_p W v_\text{D}^\mu f
   = \FMs^{-2}(
    \Tt^{\lambda}_{\nu} \partial_\lambda\FMd^{\mu\nu}
  - M 
  \FMd^{\mu\nu}\partial_\nu\FMs
)
\end{equation}
is perpendicular to the magnetic field ($\tilde\Delta^\mu_\nu \JD ^\nu=0$). We have used Eqs.~(\ref{eq:vD-dB}),  (\ref{eq:Tmunu}), and (\ref{eq:Force}) to obtain Eq.~\eqref{eq:JD-Tmunu}.

Because  of derivatives in Eq.~\eqref{eq:JD-Tmunu}, $\JD^\mu\ll\Jt^\mu$.  Therefore, to leading order, current conservation can be written as
Eq.~(\ref{eq:dJt=0}).  However, the drift current contribution cannot be ignored in Eq.~(\ref{eq:Force}) for the force, since the leading-order contribution vanishes: $B_{\nu\lambda}\Jt^\lambda=0$. 

Using Eq.~(\ref{eq:JD-Tmunu}) in Eq.~\eqref{eq:Force}, as well as the property $B^{\lambda\nu}\tilde T^\mu_{~\nu}=0$,
the energy-momentum equation in~\eqref{eq:dJT=0F}, i.e., $\partial_\mu T^\mu_{~\nu}=F_\nu$, can be rewritten as
\begin{equation}
  \label{eq:dtTmunu-F}
  \tilde\Delta_\nu^\lambda \partial_\mu \Tt^\mu_\lambda = E_{\nu\mu}\Jt^\mu
  - M\tilde\Delta_\nu^\lambda\partial_\lambda\FMs .
\end{equation}

This equation can be further simplified if one takes into account the transverse (to magnetic field) part of the stress tensor
\begin{equation}
  \label{eq:Tmunu-all}
  T^{\mu\nu} \equiv \Tt^{\mu\nu} + T_\perp^{\mu\nu},
  \quad \mbox{where}\quad T_\perp^{\mu\nu}=M\FMs \Delta^\mu_\nu\,.
\end{equation}
Then, Eq.~(\ref{eq:dtTmunu-F}) takes a simper form of Eq.~(\ref{eq:projected-dmuTmunu})~\cite{Identities}. 

Physically, the contribution $T_\perp^{\mu\nu}$ in Eq.~(\ref{eq:Tmunu-all}) comes from the fast transverse motion of the particles absent in Eq.~\eqref{eq:Tmunu}.  
 Indeed,
\begin{equation}
  \label{eq:Tmunu-ptilde}
  T^{\mu\nu}_\perp \equiv
  \int_p W\alpha\, \overline{p_\perp^\mu p_\perp^\nu} f .
\end{equation}
While $\overline{ p_\perp^\mu}=0$, $\overline{p_\perp^\mu
p_\perp^\nu} = -\Delta^{\mu\nu}p_\perp^2/2$ is nonzero.
Using Eq.~\eqref{eq:J-adiabatic} and the definition of $M$ in Eq.~\eqref{eq:Force} we obtain Eq.~\eqref{eq:Tmunu-all}.

We can also calculate the divergence of the full energy-momentum tensor in Eq.~\eqref{eq:Tmunu-all} using Eqs.~\eqref{eq:dJT=0F} and~\eqref{eq:Force}:
\begin{equation}
\label{eq:dT=FJdM} 
    \partial_\mu T^\mu_{~\nu} = F_{\nu\mu} J^\mu,
    \mbox{ with } J^\mu\equiv    \Jxdot^\mu + \partial_\lambda M^{\lambda\mu}
 \, ,
\end{equation}
where $M^{\lambda\mu}=M \FMd^{\lambda\mu}/\FMs$ is the magnetization density due to the fast transverse motion of the particles.  The full current $J^\mu$ in Eq.~(\ref{eq:dT=FJdM}) is the sum of the transport current $N^\mu=\tilde N^\mu+N_\text{D}^\mu$ and the ``magnetization current'' $\d_\lambda M^{\lambda\mu}$.  The substitution of $\JD $ from Eq.~\eqref{eq:JD-Tmunu} converts Eq.~\eqref{eq:dT=FJdM} into Eq.~\eqref{eq:projected-dmuTmunu}; i.e., only two out of four equations remain independent.

\color{black}

\emph{Constitutive relations}---The conservation laws~(\ref{eq:dJdTEJ}) form a system of three independent equations of guiding-center hydrodynamics.
All components of $\Jt^\mu$ and $T^{\mu}_{~\nu}$, thus, must be expressed through three independent variables via the constitutive relations.

At the ideal hydrodynamic level, the constitutive relations are obtained by substituting the equilibrium distribution functions $f_\text{eq}$ into the equation for the currents. This function $f_\text{eq}$ satisfies $\mathcal C[f_{\rm eq}]=0$ via detailed balance, which requires it to be 
the Fermi-Dirac or Bose-Einstein distribution function
$    f_{\rm eq}(g) = (e^g \mp 1)^{-1}$
with $g$ being a linear combination of the additive quantum numbers conserved in collisions, i.e., of 4-momentum and charge: $g=\beta u\cdot p -\alpha$.  The coefficients $\alpha$ and $\beta u^\mu$ are related to the temperature and chemical potential via $T=1/\beta$ and $\mu=\alpha/\beta$. The unit 4-vector $u^\mu$ can be identified with the frame in which the total 3-momentum vanishes.
Using Eqs.~\eqref{eq:Jpar-f-app} and \eqref{eq:Tmunu}, and performing a calculation in the frame $u=(1,\bm 0)$, where $g=\beta p_0-\alpha$, $p_0=\sqrt{p_z^2+\mt^2}$, 
and then boosting along the direction of the magnetic field, one 
obtains Eq.~(\ref{eq:JT=})
with $n$, $\epsilon$, $P$, and $P_\perp$ given by
the corresponding quantities in a one-dimensional gas, multiplied by the Landau-level density:
\begin{equation}\label{eq:nePP-feq}
  \{n, \epsilon, P, P_\perp\} = \frac{B_*}{2\pi} \!\int\! \frac{dp_z}{2\pi}\, f_\text{eq}(g)
  \biggl\{ 1, p_0, \frac{p_z^2}{p_0}, 
  \frac{JB_*}{2p_0} \biggr\}.
\end{equation}
In particular, $P_\perp=P-\FMs dP/d\FMs=-MB_*$.

{\it Second law of thermodynamics}---Ideal hydrodynamics must conserve entropy. We now show, without relying on kinetic description, i.e., without Eq.~\eqref{eq:nePP-feq}, that this requirement
implies relationships \eqref{thermo-rels} between the coefficients $n$, $\epsilon$, $P$ and $P_\perp$. 

Let $s$ be the entropy density and $su^\mu$ the entropy current.
Combining the conservation equations \eqref{eq:dJdTEJ}  with $\partial\!\cdot\! (su)=0$, we can write
\begin{equation}\label{eq:dsu}
    \partial\!\cdot\! (su) + \alpha\partial\cdot\!\tilde J
    -\beta u^\nu
    \left(\tilde\Delta_\nu^\lambda
\partial_\mu\Tt^\mu_\lambda
- E_{\nu\lambda}J^\lambda
    \right)=0 .
\end{equation}
Substituting Eqs.~\eqref{eq:JT=} into \eqref{eq:dsu} and demanding that terms proportional to $\partial\!\cdot\! u$ vanish independently from terms with $u\cdot\partial$ derivatives, we find two relations \cite{Identities}:
\begin{subequations}
\begin{align}
  & s + \alpha n - \beta (\epsilon+P) = 0, \label{eq:snep}\\
  & ds + \alpha dn -\beta d\epsilon - \beta(P-P_\perp)d\ln B_* = 0.\label{eq:dsdnde}
\end{align}
\end{subequations}
Differentiating Eq.~(\ref{eq:snep}) and comparing the result with Eq.~(\ref{eq:dsdnde}), one obtains Eqs.~(\ref{thermo-rels}).
In the kinetic theory, all relationships can be directly verified by taking derivatives  of $P$ in Eq.~\eqref{eq:nePP-feq}.

{\it Beyond kinetic theory}---%
Provided the frame velocity~$u^\mu$ is defined via $B_{\mu\nu}u^\nu=0$, $\tilde\Delta^\mu_\lambda T^\lambda_{~\nu}u^\nu=\epsilon u^\mu$,
the constitutive relations \eqref{eq:JT=} are the most general covariant relations at zeroth order in derivatives.
Therefore, the ideal hydrodynamic equations \eqref{eq:dJdTEJ} and \eqref{eq:JT=} and second law constraints (\ref{thermo-rels}) could be valid in the magnetized strongly coupled quark-gluon plasma, where kinetic theory fails.

Moreover, the hydrodynamic theory just derived should be valid in general for any conducting plasma at sufficiently long length and time scales, even when the guiding-center picture is not valid. 
In fact, Eq.~(\ref{eq:T=}) has been found previously in a related context in Ref.~\cite{Hattori:2022hyo} without using the drift approximation.

The constraint $\FMu_{\mu\nu}u^\nu=0$ comes from finite conductivity $\sigma$ of the plasma, which damps motion in directions perpendicular to the magnetic field 
at the rate 
$\Gamma_\perp=\sigma B^2/(\epsilon+P)$, which is finite even at zero wave number.  For small $B$ there is an intermediate range of timescales between the mean free time and $1/\Gamma_\perp$ where an extended hydrodynamic description with an unconstrained fluid velocity $u^\mu$~\cite{Hernandez:2017mch} is possible. This is an example of the Hydro+ regime~\cite{Stephanov:2017ghc}, where the components of velocity perpendicular to the magnetic field play the role of additional nonhydrodynamic, but slow, modes. In the true hydrodynamic regime, $u^\mu$ becomes constrained.

For a relativistic electron-positron plasma, $\sigma\sim T/e^2$.
The intermediate Hydro+ regime exists only for $B< e^3T^2$.  Note that $B\gg e^3T$ is exactly the condition when the cyclotron radius is smaller than the mean free path and drift approximation is also applicable.


The fact that there are two, not four, conservation laws for 4-momentum is a consequence of a symmetry:
A crossed field $B_{\mu\nu}$ is still invariant under a two-parameter family of diffeomorphisms $x^\mu\to x^\mu+\xi^\mu$, such that $\xi^\mu\FMu_{\mu\nu}=0$. This symmetry implies the 
conservation equations \eqref{eq:projected-dmuTmunu}.
One can also derive the constitutive relations~(\ref{eq:JT=}) from the partition function of the system in external metric and gauge fields~\cite{SM}.

Moreover, one can write down an action invariant under the constrained diffeomorphisms and encoding the isentropic solutions to our hydrodynamics.  The action is a generalization of the one previously considered in Ref.~\cite{Gralla:2018bvg}, and is related to the axion formulation of force-free electrodynamics~\cite{Thompson:1998ss,SM}.

\emph{Conclusion}---In this Letter we constructed a manifestly relativistically covariant description of the guiding-center motion based on an action principle.  This allows one to formulate the kinetic theory and write down the ideal hydrodynamic equations of a fluid of charged particles in a background of high magnetic field.  We expect our equation to be especially useful for relativistic particles with quantized transverse motion.  

Further work is needed to incorporate collisions into the guiding-center kinetic equation.  In real astrophysical problems, one needs to have in mind that in addition to the electrons, there are also ions.  From the kinetic equation with collisions term, one can determine the structure of the dissipative terms in the hydrodynamic equations and compute the kinetic coefficients~\cite{Braginskii:1965}.  Some of these coefficients are also quantities of great interest in two-dimensional physics: Hall conductivity, odd viscosity~\cite{Avron:1995fg,Avron:1998}, thermoelectric~\cite{Cooper:1997}, and thermal Hall coefficient. To determine the behavior of these coefficients in quantizing magnetic fields, one needs to understand the collisions between guiding centers.


Finally, our system of relativistically invariant equations for the guiding-center motion can be coupled to Maxwell fields via the full electromagnetic current given in Eq.~\eqref{eq:dT=FJdM}, which includes the drift current \eqref{eq:JD-Tmunu} and the magnetization current. These ingredients, together with Eq.~\eqref{eq:W} for the invariant phase-space volume, should open a new route for numerical simulations of relativistic astrophysical plasmas using the particle-in-cell or other methods~\cite{Hockney-Eastwood-book,Birdsall-Langdon-book}. 

\emph{Acknowledgement}---The authors are indebted to Jingyuan Chen, Masaru Hongo, Sanjay Reddy, Anatoly Spitkovsky, 
 and Eliot Quataert for discussions.  The authors thank the Yukawa Institute for Theoretical Physics at Kyoto University and RIKEN iTHEMS, where part of this work was completed during the workshop (YITP-T-23-05) on ``Condensed Matter Physics of QCD 2024.''  This work is supported, in part, by the U.S.\ DOE Grants No.\ DE-FG02-13ER41958 and No.\ DE-FG02-01ER41195, by the Simons Collaboration on Ultra-Quantum Matter, which is a grant from the Simons Foundation (No.\ 651440, D.T.S.), and by the United-States-Israel Binational
Science Foundation (BSF) (Grant No. 2022110).

\bibliography{guiding-center}{}

\begin{thebibliography}{46}
\bibitem [{Ide()}]{Identities}%
  \BibitemOpen
  \href@noop {} {}\bibinfo {note} {The following identities valid up to
  corrections suppressed by $\epsilon_*$ are useful:
  $\tilde\Delta^\lambda_\nu\d_\mu\Delta^\mu_\lambda=\tilde\Delta^\lambda_\nu\d_\nu
  \ln B_*$ and $u^\lambda\d_\mu \Delta^\mu_\lambda= -u^\lambda\d_\mu \tilde
  \Delta^\mu_\lambda= (u\cdot\d)\ln B_*$.}\BibitemShut {Stop}%
\end{thebibliography}


\onecolumngrid
\begin{center}
\textbf{\\
\large End Matter
\smallskip
}
\end{center}
\twocolumngrid

\setcounter{equation}{0}
\renewcommand{\theequation}{A\arabic{equation}}

\emph{Appendix: Guiding-center dynamics in curved spacetime}---%
In curved spacetime, Eq.~(\ref{eq:Lagrangian}) becomes
\begin{equation}\label{eq:Lagrangian-curved}
  L = -p_\mu\dot x^\mu - A_\mu\dot x^\mu  + \frac\alpha 2 (g^{\mu\nu} p_\mu p_\nu -\tilde m^2) + \lambda_\alpha \FMd^{\alpha\nu} p_\nu .
\end{equation}
The equations of motion following from this Lagrangian are the general coordinate invariant generalization of Eqs.~\eqref{eq:eoms}:
\begin{subequations}\label{eq:eoms-curved}
\begin{align}
  \dot x^\mu &= \alpha g^{\mu\nu}p_\nu +\lambda_\alpha \FMd^{\alpha\mu} 
  \equiv \alpha p^\mu
  + v_\mathrm{D}^\mu ,\label{dotx-curved} \\
  \dot p_\mu - \dot x^\nu \Gamma^\alpha_{\nu\mu}p_\alpha &= F_{\mu\nu}\dot x^\nu + \frac \alpha2 J
\nabla_\mu \FMs - \lambda_\alpha p_\beta \nabla_{\!\mu}\FMd^{\alpha\beta}. \label{eq:pdot-curved}
 \end{align}     
\end{subequations}
In addition, one has two constraints $p^2-\tilde m^2=0$ and $B^{\alpha\nu}p_\nu=0$.
Substituting Eq.~\eqref{dotx-curved} into Eq.~\eqref{eq:pdot-curved} and following the same steps that led to Eq.~\eqref{eq:vD-dB}, we find
\begin{equation}\label{eq:vD-dB-curved}
  v^\mu_\mathrm{D} = \alpha\FMs^{-2}\!\left(
    p_\nu (p \cdot \nabla) \FMd^{\mu\nu} +
    \frac{J}{2} \FMd^{\mu\nu} \nabla_\nu \FMs
    \right).  
\end{equation}

For a nonrelativistic particle in a constant magnetic field and a weak gravitational field ($g_{00}=1+2\Phi$, $\Phi\ll1$), Eq.~(\ref{eq:vD-dB-curved}) gives the standard gravitational drift velocity $-m B^{-2} \bm{\nabla}\Phi\times \bm{B}$~\cite{Northrop:1961}.  It is interesting to note that one and the same term in Eq.~(\ref{eq:vD-dB-curved}) is responsible for the magnetic field curvature drift and for the  gravitational drift.  This is a manifestation of Einstein's equivalence principle.
For example, consider a particle moving along a circular magnetic field line. In the frame rotating with the particle there is a centripetal force which causes the same gravitational drift as the curvature drift in the original frame. This argument is a rigorous version of a textbook derivation of the curvature drift~\cite{Jackson-EM}.

The invariant phase-space volume is given by
\begin{equation}
  \label{eq:W-curved}
    W = \frac{\FMs+B_*^{-1} p_\mu\nabla_\nu\FMd^{\mu\nu}}{2\pi^2\alpha} \theta(p_0)
  \delta(p^2-\tilde m^2)\delta^2(e_i^\mu p_\mu),
\end{equation}
where $e_i^\mu$, $i=1,2$, are two orthonormal basis vectors in the plane perpendicular to the magnetic field: $e_i^\mu e_i^\nu=\Delta^{\mu\nu}$, $g_{\mu\nu} e_i^\mu e^\nu_j=\delta_{ij}$.

Similarly, other equations in this Letter, 
such as the kinetic-theory expressions for the current and stress tensor as well as the hydrodynamic equations, generalize from their flat-spacetime counterparts 
via replacement of the momentum space integral $\int_p$ by its general-coordinate invariant version 
$(-g)^{-1/2}\!\int\!dp_0\, dp_1\, dp_2\, dp_3$, 
derivatives $\d_\mu$ by covariant derivatives $\nabla_\mu$, and $\dot p_\mu$ by the general coordinate covariant $\dot p_\mu - \dot x^\nu \Gamma^\alpha_{\nu\mu}p_\alpha$.

\clearpage
\onecolumngrid

\begin{center}
        \textbf{\large --- Supplemental Material ---\\ $~$ \\
        Relativistic Guiding-Center Motion:
        Action Principle, Kinetic Theory, and Hydrodynamics}\\
        \medskip
        \text{Dam Thanh Son and Mikhail Stephanov}
\end{center}
\setcounter{equation}{0}
\setcounter{figure}{0}
\setcounter{table}{0}
\setcounter{page}{1}
\setcounter{section}{0}
\makeatletter
\renewcommand{\thesection}{S\arabic{section}}
\renewcommand{\theequation}{S\arabic{equation}}
\renewcommand{\thefigure}{S\arabic{figure}}
\renewcommand{\bibnumfmt}[1]{[S#1]}

\section{Coordinate-space guiding-center action}

The Lagrangian~(\ref{eq:Lagrangian}) is a quadratic polynomial of $p$.  Integrating out $p$, i.e., using Eq.~\eqref{dotx}, we find
\begin{equation}\label{eq:Lagrangian-xlambda}
  L = - \frac1{2\alpha} (\dot x^\mu - \lambda_\nu B^{\nu\mu})^2 - 
  A_\mu \dot x^\mu - \frac\alpha2 \tilde m^2 .
\end{equation}
Next we integrate out $\lambda$.  Varying the action with respect to $\lambda$, we find the equation
\begin{equation}
   B_{\mu\nu}(\dot x^\nu - \lambda_\alpha B^{\alpha\nu}) = 0.
\end{equation}
The solution to this equation is
\begin{equation}
  \lambda_\alpha = \frac{B_{\alpha\beta}\dot x^\beta}{B_*^2} \,.
\end{equation}
Substituting that into Eq.~(\ref{eq:Lagrangian-xlambda}) we find
\begin{equation}\label{eq:L-x-alpha}
  L = -\frac1{2\alpha} \tilde\Delta_{\mu\nu}\dot x^\mu\dot x^\nu - A_\mu \dot x^\mu - \frac{\alpha}2 \tilde m^2 .
\end{equation}
where $\tilde\Delta_{\mu\nu}$ is defined in Eq.~(\ref{eq:projectors}).  If $\tilde m\neq0$, then one can further integrate out $\alpha$.  In the saddle-point approximation, the result reads 
\begin{equation}\label{eq:L-x}
  L = -\tilde m \sqrt{\tilde\Delta_{\mu\nu}\dot x^\mu\dot x^\nu} - A_\mu \dot x^\mu .
\end{equation}
Compared to the Lagrangian of a particle, the guiding-center Lagrangians~(\ref{eq:L-x-alpha}) and (\ref{eq:L-x}) involve an additional projector~$\tilde\Delta_{\mu\nu}$. The projector makes sure that only the motion along the magnetic field carries the usual kinetic energy, while the energy of the fast cyclotron motion is included in the rest mass $\mt$ of the guiding-center ``particle.''

\section{Comparison with Vandervoort equation~\cite{Vandervoort:1960}}

Here we show that our equation for the motion of the guiding center reproduces exactly every term in the equation previously derived by Vandervoort in Ref.~\cite{Vandervoort:1960} (see also Ref.~\cite{Northrop-book}).

We shall choose the ``worldline time'' to be equal to the coordinate time in a given (laboratory) frame. This means $x^0=\tau$, and therefore $\alpha=1/p_0$ up to terms suppressed by the small parameter $a/L$ (since $v_D\sim a/R$).
The components of the 4-momentum satisfying $p^2 =\tilde m^2$ can then be written as $p^\mu=\tilde m\gammat(1,\bm v)$, where $\gammat=1/\sqrt{1-\bm v^2}$, as usual. The spatial components of the constraint $\FMd^{\mu\nu}p_\nu=0$ read
\begin{equation}
  \label{eq:epsilon-EBv-constraint}
  \bm E' + \bm v\times\bm B' =0,
\end{equation}
where, according to Eq.~\eqref{eq:BEvsF}, up to $\mathcal O(\epsilon_*^2)$,
\begin{equation}
  \label{eq:EB-prime}
  \bm E'=\bm
  E - \epsilon_*\bm B;\quad
  \bm B'=\bm B +\epsilon_*\bm E\,.
\end{equation}
The general solution of Eq.~(\ref{eq:epsilon-EBv-constraint})
is given by
\begin{equation}
  \label{eq:constraint-solution-v}
  \bm v = v_\parallel\frac{\bm B'}{B'} + \frac{\bm E'\times\bm B'}{B'^2}\,,
\end{equation}
with arbitrary parameter $v_\parallel$.
Substituting Eqs.~(\ref{eq:EB-prime}) and keeping only the leading term
in $\epsilon_*$ in each of the three (almost) orthogonal
directions 
$\bm B$, $\bm E\times\bm B$ and $\bm E$, we find
\begin{equation}
  \label{eq:v=3terms}
  \bm v = v_\parallel \frac{\bm B}{B} +\frac{\bm E\times\bm B}{B^2} + 
  v_\parallel \epsilon_* \frac{\bm E}{B}\,.  
\end{equation}
The first term is the (arbitrary) velocity along the magnetic field direction. The second term is the usual drift velocity in crossed electric and magnetic fields:
\begin{equation}\label{eq:vperp}
    \bm v_\perp \equiv \frac{\bm E\times\bm B}{B^2} \,.
\end{equation}
The last term is the drift in the direction of the electric field caused by the (small) component of the electric field in the direction of the magnetic field:
\begin{equation}
  \label{eq:v-E-drift}
  \bm v_\perp^* \equiv v_\parallel \epsilon_* \frac{\bm E}{B} = \frac{v_\parallel
    E_\parallel\bm E}
    {\FMs^2}
    \,,
\end{equation}
where we used $\FEs\FMs=BE_\parallel$ to express
$\epsilon_*=E_\parallel B/\FMs^2$.

According to Eq.~\eqref{dotx}, total drift in the direction perpendicular to $\bm B$ is given by the sum of $\bm v_\perp$ and $\bm v_\perp^*$, which occur even if the fields are uniform, and $\bm v_\text{D}$ which occurs only in non-uniform or time-dependent fields. The latter drift is given by the spatial components of vector $v_\text{D}^\mu$ in Eq.~\eqref{eq:vD-dB}:
\begin{equation}\label{eq:vd-3}
    \bm v_\text{D} =
    \FMs^{-2}\left[
    \mt\gammat\left(
    \bm v\times\frac{d\bm B}{dt} +
    \frac{d\bm E}{dt}
        \right)
    +\frac{J}{2\mt\gammat}
        \left(
        \bm E\partial_t
        +\bm B\times\bm\nabla
            \right)\FMs
    \right],
\end{equation}
where $d/dt\equiv\partial_t +\bm v\cdot\bm\nabla$.

Following the standard practice, we shall introduce a unit vector in the direction of the magnetic field $\bm \unitb \equiv\bm B/B$. 
Using $\bm E=\bm B\times\bm v_\perp$, one can show that every term in Eq.~\eqref{eq:vd-3} is perpendicular to $\unitb$. In particular,
\begin{equation}\label{eq:vdBdt}
    \bm v\times\frac{d\bm B}{dt} +
    \frac{d\bm E}{dt} =
    B\unitb\times\left(
    v_\parallel \frac{d\unitb}{dt}+\frac{d\bm v_\perp}{dt}
    \right),
\end{equation}
where we used Eq.~\eqref{eq:v=3terms} and neglected $\mathcal O(\epsilon_*)$ term.
Putting together Eqs.~\eqref{eq:vd-3}, \eqref{eq:vdBdt}, \eqref{eq:vperp}, and \eqref{eq:v-E-drift}, we obtain, for the total drift velocity perpendicular to the magnetic field, the following expression:
\begin{equation}\label{eq:Rdot}
    \dot{\bm R}_\perp
    =\bm v_\perp + \bm v_\perp^* + \bm v_\text{D}
    = \frac{B}{\FMs^2}\unitb\times \left\{
    -\frac{\FMs^2}{B^2}{\bm E}
    + v_\parallel E_\parallel\bm v_\perp
    + \mt\gammat\left(
    v_\parallel \frac{d\unitb}{dt}+\frac{d\bm v_\perp}{dt}
    \right) + \frac{J}{2\mt\gammat}
    \left(
    \bm v_\perp\partial_t 
    + \bm\nabla
    \right)\FMs
    \right\} .
\end{equation}
This result coincides with Eq.~(243) in Ref.~\cite{Vandervoort:1960} in the form presented in Eq.~(1.76) of Ref.~\cite{Northrop-book} with the identification $\unitb\to\bm e_1$, $\bm v_\perp\to\bm u_E$ and $\FMs\to\sqrt{B^2-E_\perp^2}$.  Note that $\mt\gammat$ in Eq.~\eqref{eq:Rdot} is the total energy of the particle (including the energy of its motion around the guiding center), which in  Ref.~\cite{Northrop-book} is denoted by $m_0\gamma$.

\section{Invariant phase-space volume}

Substituting Eq.~\eqref{eq:JM} into $\partial_MJ^M=0$, we find
\begin{equation}
  \label{eq:dJ}
 0 = \partial_MJ^M = 
    \frac{\partial}{\partial x}\cdot(\dot x W)
    + \frac{\partial}{\partial p}\cdot(\dot p W)
   = 
 \alpha W\left(\frac{\partial}{\partial x}
    \cdot\!\left(\frac{\dot x}{\alpha}\right)
    + \frac{\partial}{\partial p}
    \cdot\!\left(\frac{\dot p}{\alpha}\right)
  \right)
  + \frac{1}{\alpha}\mathcal L[\alpha W],
\end{equation}
where the Liouville operator $\mathcal L$ was defined in Eq.~\eqref{eq:L=C}, and $\dot x$ and $\dot p$ are given in Eqs.~(\ref{eq:eoms}) and (\ref{eq:vD-dB}), taken to be valid in the whole 8-dimensional phase space.  We search for a solution with support 
on the intersection of the positive energy $p^0>0$ mass-shell hyperboloid $p^2=\mt^2$ and the hyperplane $B^{\mu\nu}p_\nu=0$, i.e.,
\begin{equation}
  \label{eq:Wtilde}
  W = \tilde W \,\theta(p^0)\delta(p^2-\mt^2)\delta^2(\FMd^{\mu\nu}p_\nu).
\end{equation}
The constraints $p^2-\tilde m^2=0$ and $\FMd^{\mu\nu}p_\nu=0$ are maintained by the equations of motion: $\mathcal L [p^2-\mt^2]=\mathcal L[\FMd^{\mu\nu}p_\nu]=0$.  In order to find the action of the Liouville operator on $\delta(p^2-\mt^2)\delta^2(\FMd^{\mu\nu}p_\nu)$, one needs to find how the arguments of these delta function evolve outside the constraint hypersurface.  We have
\begin{equation}
  \label{eq:L(p^2)}
  \mathcal L [p^2-\mt^2] = \alpha p^\mu(-\partial_\mu \mt^2)
  +\left(\alpha F_{\mu\nu} p^\nu + \FMu_{\mu\nu}v_\text{D}^\nu +
  \frac{\alpha}{2}\partial_\mu \mt^2 \right) (2p^\mu) = -2 v_\text{D}^\mu (B_{\mu\nu} p^\nu),
\end{equation}
and
\begin{multline}
  \label{eq:L(Bp)}
  \mathcal L[\FMd^{\mu\nu}p_\nu] =(\dot x\cdot\partial)\FMd^{\mu\nu}p_\nu
  + \dot p_\nu \FMd^{\mu\nu} =
  \alpha (p\cdot\d)B_{\mu\nu} p^\nu 
  + \left(F_{\nu\lambda} (\alpha p^\lambda + v_\text{D}^\lambda) + \frac\alpha 2 J\d_\nu B_* \right)B^{\mu\nu}\\
  = \alpha (p\cdot\d)B_{\mu\nu} p^\nu - B_*^2 \Delta^\mu_\lambda (\alpha p^\lambda + v_\text{D}^\lambda) + \frac\alpha 2 J\d_\nu B_* B^{\mu\nu} , 
\end{multline} 
Substituting the expression for the drift velocity, Eq.~(\ref{eq:vD-dB}), we find
\begin{equation}\label{eq:LBp}
  \mathcal L[\FMd^{\mu\nu}p_\nu] = \alpha \FMd^\mu_{~\lambda} (\FMu^{\lambda\nu}p_\nu).
\end{equation}
Equations~(\ref{eq:L(p^2)}) and (\ref{eq:LBp}) then imply
(since, for each of the constrained quantities $C\in\{p^2-\mt^2,B^{\mu\nu}p_\nu\}=0$, $\mathcal L[\delta[C]]=-(\partial\mathcal L[C]/\partial C)\delta[C]=0$),
\begin{equation}
  \mathcal L(\delta(p^2-\tilde m^2)\delta^2(B_{\mu\nu} p^\nu))=0.
\end{equation}
Thus $\tilde W$ obeys the same equation \eqref{eq:dJ} as $W$, i.e.,
\begin{equation}\label{eq:LW}
 \alpha \tilde W\left(\frac{\partial}{\partial x}
    \cdot\!\left(\frac{\dot x}{\alpha}\right)
    + \frac{\partial}{\partial p}
    \cdot\!\left(\frac{\dot p}{\alpha}\right)
  \right)
  + \frac{1}{\alpha}\mathcal L[\alpha\tilde W]=0.
\end{equation}

While $\d_x\cdot (\dot x/\alpha) = \d\cdot v_\text{D}=\mathcal O(\d^2)$ and can be neglected when calculating $\tilde W$ to order $\mathcal O(\partial^0)$,
\begin{multline}
  \label{eq:dpdotp}
  \frac{\partial}{\partial p}\cdot\!\left(\frac{\dot p}{\alpha}\right)
  = \FMu_{\mu\nu}\frac{\partial}{\partial p_\mu}
  \left(\frac{v_\text{D}^\nu}{\alpha}\right)
  = \FMs^{-2}\FMu_{\mu\nu}\frac{\partial}{\partial p_\mu}  \left((p\cdot\partial)\FMd^{\nu\lambda}p_\lambda \right)
= \FMs^{-2}\FMu_{\mu\nu} \left(
    (p\cdot\partial)\FMd^{\nu\mu} + \partial^\mu\FMd^{\nu\lambda}p_\lambda
  \right)\\
  = -\frac32\FMs^{-2} \FMu_{\mu\nu}(p\cdot\partial)\FMd^{\mu\nu}
  = -3(p\cdot\partial)\ln\FMs\,,
\end{multline}
where we have used Eq.~\eqref{eq:pdot} and the Bianchi identity
$\partial^{[\mu}\FMd^{\nu]\lambda}=-\frac12\partial^\lambda
\FMd^{\mu\nu} + \mathcal O(\partial \FEs)$.  Substituting Eq.~\eqref{eq:dpdotp} into Eq.~\eqref{eq:LW} we find
\begin{equation}
  \label{eq:tildeW}
  \tilde W = \tilde c\, \frac{\FMs^3}{\alpha}
  +\mathcal O(\partial),
\end{equation}
where $\tilde c$ is a numerical coefficient. 

To find $\tilde c$ it is sufficient to consider the case of a constant magnetic field pointing along the $z$ direction.  Choosing the affine parameter along the particle trajectory to be $x^0$, we have, from Eq.~\eqref{dotx}, $\alpha=1/p^0$ and thus, substituting Eqs.\eqref{eq:Wtilde} and \eqref{eq:tildeW} into Eq.~\eqref{eq:dGamma},
\begin{equation}
  d^7\Gamma = \tilde c p_0 B_*^3 \theta(p^0) \delta(p^2-\tilde m^2)\delta(B_*p_x)\delta(B_*p_y) d^3x d^4p
  = \frac {\tilde c}2 B_* \delta(p_0 - \sqrt{p_z^2+\tilde m^2})
    \delta(p_x)\delta(p_y) d^3x d^4p .
\end{equation}
Integrating over $p_0$, $p_x$, and $p_y$ we find 
\begin{equation}
  d^4\Gamma=\int_{p_0,p_x,p_y}
  d^7\Gamma = \frac {\tilde c}2 B_* dp_z d^3x .
\end{equation}
This should be compared with the density of states on a Landau level, which is given by
\begin{equation}
  d^4\Gamma=\frac{dp_z\,dz}{2\pi} \frac{B_*}{2\pi} dx\, dy.
\end{equation}
Therefore, $\tilde c=1/(2\pi^2)$.  Substituting that into Eq.~\eqref{eq:tildeW}, and using $\delta^2(B^{\mu\nu}p_\nu)=B_*^{-2}\delta^2(\Delta^{\mu\nu}p_\nu)$ in Eq.~\eqref{eq:Wtilde}, we obtain Eq.~\eqref{eq:W} up to $\mathcal O(\partial)$ terms.

There is another way to find the invariant phase-space volume. This method is, in a sense, more systematic and allows one to capture the next-order term in the gradients of the magnetic field in Eq.~\eqref{eq:tildeW}.  We introduce the vierbein $(e^0_\mu, e^1_\mu, e^2_\mu, e^3_\mu)$ so that $e^0_\mu$ and $e^3_\mu$ are the basis vectors in the 1+1d space of momenta of particles constrained to move along the magnetic field: 
\begin{subequations}
\begin{align}
   & B^{\mu\nu} e^a_\mu = 0, \qquad a=0,3, \\
   & g^{\mu\nu} e^a_\mu e^b_\nu = \eta^{ab}, \qquad \eta^{00}= -\eta^{33}=1, ~ \eta^{03}=0,\\
   & e^a_\mu e^a_\nu = \tilde\Delta_{\mu\nu}, 
\end{align}
\end{subequations}
while $e^1_\mu$ and $e^2_\mu$ are the two basis vectors perpendicular to the magnetic field, 
\begin{subequations}
\begin{align}
   & \tilde B^{\mu\nu} e^i_\mu = 0, \qquad i=1,2, \\
   & g^{\mu\nu} e^i_\mu e^j_\nu = -\delta^{ij}, \\
   & e^i_\mu e^i_\nu = \Delta_{\mu\nu} .
\end{align}
\end{subequations}

We expand $p_\mu$ in terms of the vierbein: $p_\mu=e^a_\mu p_a+ e^i_\mu p_i$.  The constraint $B^{\mu\nu} p_\nu =0$ can now be solved: it implies that $p_i=0$. The Lagrangian is
then 
\begin{equation}\label{eq:L6d}
  L = - p_a e^a_\mu \dot x^\mu - A_\mu \dot x^\mu + \frac\alpha2 (p^2 - \tilde m^2),
\end{equation}
and the equations of motions are 
\begin{subequations}
  \label{eq:4bein2}
\begin{align}
    & - e^a_\mu \dot p_a +[F_{\mu\nu} + p_a(\d_\mu e^a_\nu - \d_\nu e^a_\mu)] \dot x^\nu
  + \frac\alpha2 \d_\mu\tilde m^2 =0,\\
    & e^a_\mu\dot x^\mu - \alpha p^a=0,
\end{align}
\end{subequations}
which, as one can check, are equivalent to Eqs.~\eqref{eq:eoms}.

By combining the four coordinates $x^\mu$ and two components of the momentum $p_a$ into a six-component degree of freedom,
\begin{equation}
  \xi^{\cal M} = \begin{pmatrix} x^\mu\\ p_a \end{pmatrix},  
\end{equation}
the Lagrangian in Eq.~\eqref{eq:L6d} can be rewritten in a general form
\begin{equation}
  L = - \mathcal A_\cM(\xi) \dot\xi^\cM - H(\xi),  
\end{equation}
where 
\begin{equation}
   \mathcal A_\cM = \begin{cases} A_\mu + p_a e_\mu^a,  & \xi^\cM = x^\mu , \\
    0, & \xi^\cM = p_a ,
       \end{cases}
\end{equation}
and $H(\xi) = -\alpha (p^2-\tilde m^2)/2$. 
The equations of motion \eqref{eq:4bein2} can be written in the general form
\begin{equation}
   \mathcal F_{\cM\cN}\dot \xi^\cN + \frac{\d H}{\d \xi^\cM}=0,
\end{equation}
where
\begin{equation}
  \mathcal F_{\cM\cN} = \d_\cM A_\cN - \d_\cN A_\cM  
  = \begin{pmatrix} 
  F_{\mu\nu}+p_a(\d_\mu e^a_\nu-\d_\nu e^a_\mu) & & -e^b_\mu\vspace{6pt}\\
  e^a_\nu & & 0
  \end{pmatrix} .
\end{equation}
Defining $\mathcal F^{MN}$ as the matrix inverse of $\mathcal F_{MN}$, the velocity can be written as
\begin{equation}\label{eq:xidot}
  \dot\xi^\cM = - \mathcal F^{\cM\cN} \frac{\d H}{\d\xi^\cN}\,. 
\end{equation}
which, one can check, is equivalent to equation of motion \eqref{eq:eoms}.
Since $\mathcal F_{MN}$ is an antisymmetric metric, its inverse can be written as
\begin{equation}\label{eq:F-inverse}
   \mathcal F^{\cM\cN} = -\frac1{2 \mathop{\mathrm{Pf}}(\mathcal F)} \epsilon^{\cM\cN\cM_1\cN_1\cM_2\cN_2} \mathcal F_{\cM_1\cN_1}\mathcal F_{\cM_2\cN_2},
\end{equation}
where $\mathop{\mathrm{Pf}}(\mathcal F)$ is the Pfaffian of the matrix $\mathcal F_{MN}$,
\begin{equation}
  \mathop{\mathrm{Pf}}(\mathcal F) = \frac16 \epsilon^{\cM_1\cN_1\cM_2\cN_2\cM_3\cN_3}\mathcal F_{\cM_1\cN_1} \mathcal F_{\cM_2\cN_2}\mathcal F_{\cM_3\cN_3} .
\end{equation}

We now show that the following six-current is conserved:
\begin{equation}
   J^\cM = \mathop{\mathrm{Pf}}(\mathcal F)\, \dot\xi^\cM .
\end{equation}
Indeed, from Eqs.~(\ref{eq:xidot}) and (\ref{eq:F-inverse}) it follows that
\begin{equation}
  J^\cM = \frac12 \epsilon^{\cM\cN\cM_1\cN_1\cM_2\cN_2} \mathcal F_{\cM_1\cN_1}\mathcal F_{\cM_2\cN_2} \d_\cN H.
\end{equation}
Conservation of $J^\cM$ is now the consequence of the Bianchi identity $\d_{[\cM}\mathcal F_{\cN\cL]}=0$.

The Pfaffian of the matrix $\mathcal F$ is a Lorentz scalar and can be computed by choosing the local reference frame where the only nonzero component of $B_{\mu\nu}$ is $B_{12}=-B_{21}$ and then rewriting the result in the covariant notation.  We find
\begin{equation}
  \mathop{\mathrm{Pf}}(\mathcal F) = \frac{B^{\mu\nu}}{2B_*} \mathcal F_{\mu\nu} =  B_* + \frac{p_a}{2B_*} B^{\mu\nu}(\d_\mu e^a_\nu - \d_\nu e^a_\mu) 
  = B_* + \frac{p_\mu}{B_*} \d_\nu B^{\mu\nu}.
\end{equation}
To perform the last transformation we used $B^{\mu\nu}e^a_\nu=0$. Thus, the invariant phase-space volume is given by
\begin{equation}\label{eq:WdB}
    W = \frac{\tilde c}{\alpha}
    \left(\FMs 
    + \FMs^{-1}p_\mu\partial_\nu
    \FMd^{\mu\nu}\right)\theta(p^0)\delta(p^2-m^2)\delta^2(p_i)
   = \frac{\tilde c}{\alpha}
    \left(\FMs 
    + \FMs^{-1}p_\mu\partial_\nu
    \FMd^{\mu\nu}\right)\theta(p^0)\delta(p^2-m^2)\delta^2(\Delta^{\mu\nu}p_\nu).
\end{equation}

\section{Conservation laws from symmetries}
\label{app:symmetries}

The conservation laws at the core of the hydrodynamics are consequences of symmetries. 
We demonstrate this, as usual, by defining the energy-momentum tensor $T^{\mu\nu}$ and current $J^\mu$ as a response of the system to the perturbations of external metric $g_{\mu\nu}$ and electromagnetic vector potential $A_\mu$:
\begin{equation}\label{eq:dlnZ}
   \delta\ln Z[g_{\mu\nu},A_\mu] = - \!\int\!d^4x\, \sqrt{-g} \left( \frac12 T^{\mu\nu}\delta g_{\mu\nu} + \JZ^\mu \delta A_\mu\right) ,
\end{equation}
where $Z[g_{\mu\nu},A_\mu]$ is the generating functional. Then the gauge invariance of the generating functional $Z$ implies the conservation of current $\JZ^\mu$,
since under local gauge transformations $\delta_\alpha A_\mu=\partial_\mu\alpha$ the variation of $\ln Z$ is given by
\begin{equation}\label{eq:dlnZ-alpha}
  0=\delta_\alpha \ln Z = - \!\int\! d^4x\,\sqrt{-g}\, \JZ^\mu \delta_\alpha A_\mu
  = \int\! d^4x\, \sqrt{-g}\, \alpha \nabla_\mu \JZ^\mu\,.
\end{equation}

Similarly, the invariance 
of the generating functional $Z$
under local diffeomorphisms
\begin{align}
  \mathcal L_\xi g_{\mu\nu} &= \xi^\lambda\d_\lambda g_{\mu\nu} + g_{\lambda\nu}\d_\mu\xi^\lambda + g_{\mu\lambda} \d_\nu\xi^\lambda
  = \nabla_\mu\xi_\nu + \nabla_\nu\xi_\mu\,,
  \\
  \mathcal L_\xi A_{\mu} &= \xi^\lambda\d_\lambda A_\mu + A_\lambda\d_\mu\xi^\lambda
  = \xi^\lambda F_{\lambda\mu} + \partial_\mu(\xi^\lambda A_\lambda)\,,
  \label{eq:LxiA}
\end{align}
where $\mathcal L_\xi $ denotes Lie derivative, implies
\begin{equation}\label{eq:0=LZ}
    0 =   \mathcal L_\xi\ln Z = - \!\int\!d^4x\, \sqrt{-g} \left( \frac12 T^{\mu\nu}\mathcal L_\xi g_{\mu\nu} + \JZ^\mu \mathcal L_\xi A_\mu\right) 
       =
         \int\!d^4x\, \sqrt{-g}\, \xi^
   \lambda (\nabla_\mu T^\mu_{~\lambda} - F_{\lambda\mu}\JZ^\mu +
   A_\lambda\nabla_\mu \JZ^\mu
   )\,.
\end{equation}
Using current conservation $\nabla_\mu \JZ^\mu=0$, we arrive at $\nabla_\mu T^\mu_\lambda=F_{\lambda\mu} \JZ^\mu$. 

Non-conservation of the energy-momentum tensor is related to the presence of the external field $F_{\mu\nu}$ which violates diffeomorphism invariance, since
\begin{equation}
    \mathcal L_\xi F_{\mu\nu}
    =
    \xi^\lambda\d_\lambda F_{\mu\nu} + F_{\lambda\nu}\d_\mu\xi^\lambda + F_{\mu\lambda} \d_\nu\xi^\lambda
    = \d_\mu(\xi^\lambda F_{\lambda\nu})
    -
    \d_\nu(\xi^\lambda F_{\lambda\mu})\,
\end{equation}
is nonzero in general.
However, the diffeomorphisms which obey $\xi^\lambda F_{\lambda\mu}=0$ do not change $F_{\mu\nu}$ (in this case $\mathcal L_\xi A_\mu$ is a gauge transformation), and thus would lead to the conservation of the corresponding components ($T^\mu_{~\nu}\xi^\nu$) of the energy-momentum. In general, there are no such diffeomorphisms. However, for a crossed field $F_{\mu\nu}=\FMu_{\mu\nu}$, i.e., for $\FEs=0$, solutions of $\xi^\lambda\FMu_{\lambda\nu}=0$ do exist and define a two-dimensional hyperplane. Vectors $\xi^\lambda$ in this hyperplane can be expressed as $\xi^\lambda=\tilde\Delta^\lambda_\nu\eta^\nu$, where $\eta^\nu$ is an arbitrary 4-vector. Using such diffeomorphisms in Eq.~\eqref{eq:0=LZ} one obtains projected conservation equations 
$\tilde\Delta_\lambda^\nu\partial_\mu T^\mu_{~\nu}=0$. 
If $\FEs\neq0$, then these projected equations become projected non-conservation Eqs.
\eqref{eq:projected-dmuTmunu}.

\section{Constitutive relations from a partition function}

We can derive the constitutive relations \eqref{eq:JT=} for ideal hydrodynamics from the equilibrium generating functional (or partition function) $Z$  expressed in terms of the pressure $P$ as a function of temperature and chemical potential through $\alpha=\mu/T$, $\beta=1/T$ as well as magnetic field $B_*$:
\begin{equation}\label{eq:lnZP}
    \ln Z = \int\! d^4x\, \sqrt{-g}\, P(\alpha,\beta,\FMs)\,.
\end{equation}
The local equilibrium state of the system is characterized by vector $\beta^\mu$ such that $\mathcal L_\beta g_{\mu\nu}=\mathcal L_\beta A_\mu=0$. This vector encodes both the local temperature and the local rest frame 4-velocity $u^\mu$ through
\begin{equation}
  \beta^\mu = \beta u^\mu,
\end{equation}
so that $\beta=\sqrt{\beta^\mu\beta^\nu g_{\mu\nu}}$ and $u^\mu=\beta^\mu/\beta$, while $\alpha=-\beta^\mu A_\mu$.

Evaluating variational derivatives in Eq.~\eqref{eq:dlnZ} and keeping only the leading, i.e., zeroth order, in gradients of the magnetic field, we obtain Eqs.~(\ref{eq:JT=}).
In particular, the relation 
\begin{equation}
    \frac{\partial\FMs}{\partial g_{\mu\nu}}=-\frac12\FMs\Delta^{\mu\nu}\,,
\end{equation}
where $\FMs=(\FMu_{\mu\alpha}\FMu_{\nu\beta}g^{\mu\nu}g^{\alpha\beta}/2)^{1/2}$, explains the appearance of the $P_\perp\Delta^{\mu\nu}$ term in Eq.~\eqref{eq:T=}.

We can also obtain the total current, including the contribution of first order in the gradients of the magnetic field:
\begin{equation}\label{eq:dlnZdA}
\Jtot^\mu= -\frac{\delta\ln Z}{\delta A_\mu} = n u^\mu 
    + \partial_\lambda\left(
\frac{\partial P}{\partial\FMs}
\frac{\FMd^{\lambda\mu}}{\FMs}
    \right).
\end{equation}
This result matches the value of the total current we found in Eq.~\eqref{eq:dT=FJdM}, 
\begin{equation}\label{eq:Jtot}
    \Jtot^\mu
    =\Jxdot^\mu + \partial_\lambda(M\FMd^{\lambda\mu}/\FMs)
    =\Jt^\mu + \JD ^\mu +\partial_\lambda(M\FMd^{\lambda\mu}/\FMs),
\end{equation}
evaluated in equilibrium. To see this, we use Eq.~\eqref{eq:Jpar-f-app} for the transport current $\Jt$ along the magnetic field, where we must take into account terms first order in the gradients of the magnetic field in the invariant volume $W$ given by Eq.~\eqref{eq:WdB}. We find
\begin{equation}\label{eq:Jt-equil}
    \Jt^\mu =nu^\mu + 
    \tilde\Delta^\mu_\nu
    \partial_\lambda\left(
\frac{P}{\FMs}
\frac{\FMd^{\lambda\nu}}{\FMs}
    \right).
    \end{equation}
For the drift current, using Eq.~\eqref{eq:JD-Tmunu} and substituting Eq.~\eqref{eq:T=} we find:
\begin{equation}\label{eq:JD-equil}
    \JD^\mu =\Delta^\mu_\nu
    \partial_\lambda\left(
\frac{P}{\FMs}
\frac{\FMd^{\lambda\nu}}{\FMs}
    \right),
    \end{equation}
where we used
\begin{equation}\label{eq:MPB}
    M=-\frac{P_\perp}{\FMs}=
    \frac{\partial P}{\partial\FMs}-
    \frac{P}{\FMs}\,.
\end{equation}
Substituting Eqs.~\eqref{eq:Jt-equil},~\eqref{eq:JD-equil}, and~\eqref{eq:MPB} into Eq.~\eqref{eq:Jtot} for the total current we find that it matches the current in Eq.~\eqref{eq:dlnZdA} obtained by varying the partition function in Eq.\eqref{eq:lnZP}.

\section{Axion formulation of hydrodynamics of isentropic flows}

If one restricts oneself to isentropic flows, then from the set of three hydrodynamic variables, there remains only two: charge density and velocity.  Two first-order hydrodynamic equations is equivalent to one second-order equation, and one may wonder if there exist an action, containing fields with no more than one derivative, that yields that second-order equation.  Indeed, one can capture the whole isentropic hydrodynamics with the Lagrangian
\begin{equation}\label{axion-L}
  \mathcal L = -\epsilon\left( (\tilde N^\mu \tilde N_\mu)^{1/2}, B_* \right)
  + \frac\varphi{16\pi^2} F_{\mu\nu}\tilde F^{\mu\nu},
\end{equation}
where here, by definition $\tilde N^\mu= -\frac1{4\pi^2}\tilde F^{\mu\nu}\d_\nu\varphi$ and $\epsilon(n, B_*)$ is the function giving the dependence of the energy density on the charge density and magnetic field at a fixed $s/n$.  

The correspondence with hydrodynamics variables is established by identifying $\tilde N^\mu=n u^\mu$, making Eqs.~(\ref{eq:dJt=0}) and (\ref{eq:J=}) automatic. To leading order in $\epsilon_*\ll1$, we also have $B_{\mu\nu} u^\nu=0$.  By introducing a metric, one can check that the stress tensor has the form~(\ref{eq:T=}). 
Then, according to the discussion in Sec.~\ref{app:symmetries}, energy-momentum conservation leads to Eq.~(\ref{eq:projected-dmuTmunu}).

In the crossed-field backgrounds where $F_{\mu\nu}\tilde F^{\mu\nu}=0$, the theory~(\ref{axion-L}) has a nontrivial subsystem symmetry.  One notes that field configurations where $F\tilde F=0$ can be parameterized by two functions $X$ and $Y$,
\begin{equation}
  F_{\mu\nu} = \d_\mu X \d_\nu Y - \d_\nu Y \d_\mu X.
\end{equation}
The action~(\ref{axion-L}) is then invariant under
\begin{equation}
  \varphi \to \varphi + f(X,Y),  
\end{equation}
where $f$ is an arbitrary function of two variables.  This symmetry simply means that, in ideal hydrodynamics, particle number (and entropy) is conserved along each magnetic field line $X=\text{const}$, $Y=\text{const}$, 

One concrete example is that of electrons at zero temperature, at densities low enough that they are all on the lowest Landau level.  The equation of state in the ultra-relativistic limit (requiring $B\gg 4.4\times 10^{13}~\text{G}$) of such a cold electron gas is, 
\begin{equation}
   \epsilon(n, B_*) = \frac{\pi^2 n^2}{B_*} \,,
\end{equation}
hence the Lagrangian~(\ref{axion-L}) becomes
\begin{equation}
  \mathcal L = \frac{B_*}{16\pi^2}\tilde\Delta^{\mu\nu}\d_\mu\varphi \d_\nu\varphi +  \frac\varphi{16\pi^2} F_{\mu\nu}\tilde F_{\mu\nu}.
\end{equation}
Except for the absence of a $\cos\varphi$ term, this action has exactly the same form as the one considered in Ref.~\cite{Gralla:2018bvg}.  This theory thus describes the physics in a thin layer in the crust of magnetars with ultrahigh magnetic fields, where electrons are on a single Landau level.

Adding the Maxwell term $-\frac14 F_{\mu\nu}^2$ and discarding the $\epsilon(n)$ term, one obtains the axion formulation of force-free electrodynamics~\cite{Thompson:1998ss}.

\end{document}